# Magnetic relaxation in $La_{0.250}Pr_{0.375}Ca_{0.375}MnO_3$ with varying phase separation

I. G. Deac,[1,2] S. V. Diaz,[1,3] B. G. Kim,[4] S.-W. Cheong[4] and P. Schiffer[1*]

[1]Department of Physics and Materials Research Institute, Pennsylvania State University, University Park, PA 16802

[2]Department of Physics, Babes-Bolyai University, 3400 Cluj-Napoca, Romania

[3]Faculté de Physique, Université Joseph Fourier, Saint Martin d' Hères 38041, France

[4]Department of Physics and Astronomy, Rutgers University, Piscataway, New Jersey 08854

We have studied the magnetic relaxation properties of the phase-separated manganite compound $La_{0.250}Pr_{0.375}Ca_{0.375}MnO_3$. A series of polycrystalline samples was prepared with different sintering temperatures, resulting in a continuous variation of phase fraction between metallic (ferromagnetic) and charge-ordered phases at low temperatures. Measurements of the magnetic viscosity show a temperature and field dependence which can be correlated to the static properties. Common to all the samples, there appears to be two types of relaxation processes – at low fields associated with the reorientation of ferromagnetic domains and at higher fields associated with the transformation between ferromagnetic and non-ferromagnetic phases.

[*] *Corresponding author: schiffer@phys.psu.edu*

## I. Introduction

The perovskite manganites with composition $Ln_{1-x}R_xMnO_3$, where "Ln" is a lanthanide and "R" is an alkaline earth, are a particularly important class of materials because of the close coupling between their structural, magnetic and transport properties [1]. Recent work has suggested that the magnetoelectronic ground state of many of these materials can be inhomogeneous due to the coexistence of a ferromagnetic (FM) metallic phase and an antiferromagnetic (AFM) charge ordered (CO) phase and that this phase separation governs many of the observed physical properties of these materials [2,3,4,5,6].

The coexistence of ferromagnetic and antiferromagnetic tendencies in the manganites suggests that there should be glassy behavior associated with the frustration of the different interactions. Indeed logarithmic time dependence has been observed in the relaxation of both resistivity and magnetization after sudden changes in magnetic field [7,8,9,10,11,12,13,14,15,16,17], and there have been numerous reports of spin-glass-like behavior in these materials [18,19,20,21,22] (although the qualitative details of the behavior are not necessarily consistent with the conventional behavior of spin glasses [23]). While this sort of glassy behavior and the slow relaxation can be attributed to domains in ferromagnets, to superparamagnetism or to conventional spin glass phases associated with local disorder, another possibility in the phase separated materials is that the relative fractions of the different phases slowly adjust to changes in the external environment. This explanation is consistent with the observation that the relaxation effects seem to be especially pronounced in materials near $x = 0.50$ [8,11,15,16] where there is often a crossover between predominantly ferromagnetic conducting and antiferromagnetic charge ordered states [24] and phase separation appears to dominate the physical properties [2,25].

In this work we investigate a series of polycrystalline samples with nominal composition $(La_{0.250}Pr_{0.375})Ca_{0.375}MnO_3$ in which the relative fraction of the coexisting phases is varied, but the charge-ordering and Curie temperatures remain virtually constant. In particular, we focus on the temperature and magnetic field dependence of the magnetic relaxation after a sudden decrease in the applied magnetic field. We find that the magnetic viscosity has a complex temperature dependence which correlates well with the static properties of the material. Furthermore, we can identify two different relaxation mechanisms, at large and small applied magnetic fields, which appear to be associated with fluctuations between ferromagnetic and charge-ordered or paramagnetic phases and reorientation of ferromagnetic domains respectively.

## II. Experimental Details

Our polycrystalline samples of $La_{0.250}Pr_{0.375}Ca_{0.375}MnO_3$ were prepared from $La_2O_3$, $Pr_6O_{11}$,$CaCO_3$, and Mn metal by a citrate gel technique. Stoichiometric amounts of the starting materials were dissolved in diluted nitric acid ($HNO_3$) to which was added an excess of citric acid and ethylene glycol to make metal-complex. After all the reactants had completely dissolved, the solution was mixed and heated on a hot plate resulting in the formation of a gel. The gel was dried at 300 $^0$C , then heated to 600 $^0$C to remove the organic and to decompose the nitrates. The resulting powder was pressed into pellets, each of which was sintered at a different temperature (900, 950, 1000, 1100, 1200, and 1300 $^{0}$C). The samples we studied are thus labeled by their sintering temperature. All the samples showed single-phase x-ray patterns, and the average grain sizes (estimated by SEM) increased with increasing annealing temperature as discussed below.

The resistivity and magnetization of the samples were measured in Quantum Design PPMS cryostats and MPMS SQUID magnetometers respectively. The latter systems were also used to study the time dependence of the magnetization with the following procedure: 1) the superconducting magnet was degaussed by oscillating the magnetic field; 2) the sample was zero-field-cooled from 300 K to a base temperature $T_0$; 3) the sample was held at $T = T_0$ for a time $t_{w1} = 1000$ s; 4) the magnetic field was ramped up and maintain to a value $H_a$ for another $t_{w2} = 1000$ s; 5) the magnetic field was reduced to near zero and the magnetization was recorded for a time interval longer than 3500 s. The end field was chosen to be 0.002 T (which is slightly higher than the typical magnitude of the residual field in these cryostats), in order to avoid a reversed residual field. The applied magnetic field, $H_a$, was set to 0.05, 0.1, 0.2, 0.3, 0.5, 1, 2 and 4 T (it takes 200 s to ramp the field to 1 T) and the time-dependent measurements were performed at 2, 5, 10, 25, 50, 75, 100, 130 and 160 K. We fit the relaxation of the magnetization with the form: $M(t) = M(t=0)\left[1 - S(T, H_a)\log(\frac{t}{\tau_0})\right]$, where $S(T, H)$ is the magnetic viscosity which characterizes the relaxation process [26,27].

We found that the magnetometers were subject to a residual time dependent magnetic field of the form $H'(t) = H_0' + a\log[t]$ (presumably attributable to relaxation in the magnet or in nearby magnetic materials in the building). The residual field parameters, i.e. $H_0^{'}$ and $a$, depended on the applied field $H_a$, and were temperature independent (typical values were $a =$ 0.014 Oe and $H_0^{'}$ = 24.7 Oe for $H_a$ = 0.05 T; $a$ = 0.142 Oe and $H_0^{'}$ =3.92 Oe for $H_a$ = 4 T). The magnetic moment induced in the $La_{0.250}Pr_{0.375}Ca_{0.375}MnO_3$ samples by this field was obtained from the measured low-field susceptibility and then subtracted from the measured magnetic moment in order to obtain the actual magnetic viscosity. This correction was small, but significant when S approached zero.

## III. Results

### A. Time independent measurements

We chose to examine a composition from the series $(La_{5/8-y}Pr_y)Ca_{3/8}MnO_3$ in which the varying Pr fraction ($y$) controls the fraction of the ferromagnetic metallic phase in the ground state. Previous studies have demonstrated two-phase coexistence, involving a predominantly FM metallic state at small $y$, a predominantly charge-ordered state at large $y$, and the relative fractions changing with $y$ [28,29,30,31]. Furthermore, there is strong evidence that all of the magnetotransport properties are dominated by coexistence of the two phases, since the application of a magnetic field simply converts sample from the charge-ordered state to the ferromagnetic conducting state. Kiryukhin *et al*. [32] examined the phase diagram for y = 0.35, and they found that with decreasing temperature the compound first undergoes a transition to a predominantly charge-ordered phase at $T_{CO} \approx 200$ K and then an insulator-metal transition at $T \sim 70$ K to a mixed phase consisting of CO insulating, paramagnetic, and FM metallic regions. They also concluded that a non-charge-ordered paramagnetic phase is present below $T_{CO}$ which coexists with the charge-ordered phase, and that it (rather than the charge-ordered phase) evolves into the ferromagnetic metallic phase below $T_C$ in zero magnetic field.

Due to the strong coupling between the lattice and other physical properties, the strain associated with the grain boundaries and other type of structural defects can dramatically affect the electrical and magnetic behavior of polycrystalline manganites [33,34]. In our series of polycrystalline samples with $y = 0.375$, we find that the grain size increases monotonically with annealing temperature, as shown in figure 1. Since this composition is near the edge of the compositional transition between the metallic and insulating ground states, one might expect a dramatic effect on the resistivity from this change in grain size. Indeed, we do observe a striking difference between the samples in the temperature dependent resistivity which is plotted in figure 2 in zero applied magnetic field and in a field of 10 T. For samples 900 and 950, the resistivity increases monotonically with decreasing temperature in zero magnetic field to the lowest temperature we could measure. For the other samples, the zero-field resistivity begins to decrease at a temperature near the onset of ferromagnetism seen in the magnetization data (discussed below). All samples have a sharply suppressed resistivity in the high magnetic field, and this large magnetoresistance is presumably associated with a field-induced ferromagnetic metallic state indicated by the saturated moment observed in magnetization measurements as a function of field (see figure 4 below). The onset of charge ordering is discernable as a steep change in the resistance around $T_{CO} \sim 220$ K for all the samples, and $T_{CO}$ can be determined from a maximum in the slope of the resistivity. The electrical behavior of the samples is similar to that reported previously [28,29,35] for (La,Pr,Ca)$MnO_3$ in which the magnetotransport data could be described in terms of percolative transport through the ferromagnetic metallic regions which increased in volume with La content. The resistivity data indicate that the fraction of the metallic phase analogously increases monotonically with annealing temperature (and grain size) in the series of samples studied here and also increases with applied magnetic field at all temperatures below $T_{CO}$.

In figure 3, we show the temperature dependence of the magnetization of the samples in a 1 T magnetic field, showing a feature near $T_{CO} \sim 200$ K, and a sharp rise at lower temperatures where part of the sample becomes ferromagnetic. The Curie temperature, $T_C \sim 85$ K, can be determined from these data as the maximum in the temperature derivative. In figure 4, we show the field dependence of the magnetization, *M(H)*, after zero field cooling at various temperatures for sample 1100. This plot demonstrates paramagnetic behavior at high temperatures (*M*(*H*) is linear), and evidence for ferromagnetism at the lowest temperatures (in that there is downward curvature of *M*(*H*)). Between $T_{CO}$ and $T_C$, there is a small ferromagnetic rise in *M*(*H*) and then upward curvature -- presumably indicating the presence of a small ferromagnetic fraction which grows to encompass the entire sample in a sufficiently strong field. At the lowest temperatures *M*(*H*) exhibits a plateau at around 1 T which presumably corresponds to fields which are sufficient to align the moments of those regions which are ferromagnetic at zero field, but which are not sufficient to transform charge-ordered regions into ferromagnetism. We can estimate the volume fraction which is ferromagnetic after zero field cooling ($f_{FM}$) by taking the ratio of the plateau magnetization to the theoretically expected magnetization (4.45 μB/f.u.) for complete alignment of all the Mn moments. The values of $f_{FM}$ at 5 K for all the samples are plotted in figure 1, and it is notable that the ferromagnetic fraction at low temperature increases with annealing temperature as the samples become more conducting. Furthermore, when $f_{FM}$ exceeds the three-dimensional percolation threshold of 14.5% [36] for samples 1000-1300, the low temperature zero-field state of the samples changes from insulating to conducting. Thus we can conclude that the relative fractions of the two phases are changing across our sample series, but that the individual phases (as characterized by $T_{CO}$ and $T_C$) are unchanged. It is interesting to note that our observed dependence of $f_{FM}$ on grain size and annealing time is opposite that seen by Levy *et al.* in $La_{0.5}Ca_{0.5}MnO_3$ [37], perhaps due to the inverted order of $T_C$ and $T_{CO}$ in that material compared to $(La_{5/8-y}Pr_y)Ca_{3/8}MnO_3$. Due to the smooth variation of $f_{FM}$ in our series, we can thus examine the relaxation process as a function of the fractions of the coexisting phases, as described in the following section.

**B. Relaxation measurements**

We studied the magnetic relaxation of each of the samples in the series using the procedure described above to obtain the magnetic viscosity. We focused our studies on low temperatures (T $\leq$ 160 K) in order to examine the regions in which two-phase coexistence clearly dominates the behavior. Figure 5 shows representative profiles of the magnetic relaxation for sample 1000 with an applied field of $H_a = 0.05$ T and different temperatures (figure 5a) and with different applied magnetic fields for $T = 130$ K (figure 5b). The magnetization in each case was normalized to the corresponding value at $t = 0$ (immediately after the field was stabilized near zero) and $M(t)/M(0)$ was plotted versus log($t$). The asymptotic slopes of these curves, i.e. the long time logarithmic relaxation rates, were used to estimate the magnetic viscosity, $S = \frac{1}{M_0} \frac{dM}{d(\log(t))}$ which was then corrected for the residual field as described above.

The temperature dependence of $S$ for sample 1100 is shown in figure 6 for a range of applied magnetic fields ($H_a$). As can be seen in this figure, the temperature dependence of $S$ has a characteristic behavior in low applied fields (open symbols, $H_a \leq 0.3$ T) and a different characteristic behavior in larger fields (solid symbols, $H_a \geq 0.5$ T). The difference between the low and high field regimes can also be seen by plotting the same data (figure 7) as a function of the applied field at different temperatures. Although $S$ is small and nearly independent of the applied field at the lowest temperatures, for all higher temperatures there is a clear difference between the two field regimes. At $T = 25$ K and 50 K, $S$ has a large magnitude for small fields, but approaches zero rapidly up to $H_a \sim 0.5$ T where it saturates. For temperatures above 50 K, $S$ has weak field dependence up to $H_a \sim 0.5$ T, then it dramatically increases in magnitude for larger fields.

Since the magnetic viscosity appears to have different behavior for low and high values of the applied magnetic field $H_a$, we must conclude that there are two different relaxation processes responsible for the magnetic viscosity in this series of materials. We first discuss the small $H_a$ processes shown in figure 8, where $S$ has an almost constant small magnitude at the lowest temperatures (T $\lesssim$ 10K), has the largest magnitude just below $T_C$, and then has a small and almost temperature-independent value above $T_C$. We do not attribute the observed relaxation processes in this low field range to changes in the relative fractions of ferromagnetic and non-ferromagnetic regions because there is little magnetoresistance at such low fields (which indicates that there is little conversion from charge order to metallic conduction). There is also no consistent variation among the obtained values of $S$ between the samples in the series, even though the resistivity and equilibrium magnetization data discussed above indicate a changing ferromagnetic fraction among the samples. We attribute the relaxation in this low field regime instead to relaxation of the moment orientation of the ferromagnetic regions in the material which exist even after zero-field-cooling. This explanation is consistent with the largest viscosity being observed just around $T_C$ where the moments are more easily reoriented. We also observe that $S$ at the lowest temperatures has a larger magnitude for samples 900 and 950 than for all of the other samples. Those samples are below the percolation threshold in their ferromagnetic fraction, which presumably would have an effect on the magnetic relaxation processes due to anisotropy factors.

The temperature dependence of $S$ is quite different at the higher magnetic fields as shown in figures 9 and 10. In this field regime, $S$ is almost constant for $T < T_C$, but then increases abruptly in magnitude near $T_C$ and for higher temperatures, i.e. in the predominantly charge-ordered state. This temperature dependence is qualitatively similar to that observed by López *et al.* in $La_{0.5}Ca_{0.5}MnO_3$ after reducing the field from 5 tesla (the only magnetic field from which they studied relaxation). We attribute relaxation in this regime to the fraction of ferromagnetic volume decreasing after the removal of the magnetic field, i.e. to the relaxation of the proportions of the coexisting phases. This view is supported by the data in figure 11 where we plot $S$ for $H_a$ = 1 T as a function of sample number (where the higher numbered samples were sintered at higher temperature and have a larger ferromagnetic phase fraction as discussed above). While the data are somewhat noisy at the lowest temperatures (due to the smaller magnitude of $S$), we see that $S$ increases in magnitude significantly (by a factor of 2 or more) with increasing ferromagnetic phase fraction in this high-field regime. Since both $T_C$ and $T_{CO}$ are approximately constant for all of the samples, the change in $S$ among the samples can be attributed to the changing ferromagnetic fraction, and thus the relaxation itself to relaxation of the proportions of the coexisting phases. The increase in S we observe just above $T_C$ in this field regime is also consistent with this explanation. While for $T < T_C$ the ferromagnetic fraction remains roughly constant after removing the magnetic field, the application of a strong magnetic field to the CO phase converts some fraction of it to a ferromagnetic conducting phase (see figure 4). After reducing the applied magnetic field, the system reverts to the equilibrium predominantly charge-ordered state [28], and we are apparently observing the relaxation of that process. An interesting point in the data is the trend of increasing $S$ with the increasing sample number which correlates with an increased fraction of the ferromagnetic state within the predominantly charge-ordered matrix. In other words, the system relaxes faster when the end point has a larger fraction of the minority ferromagnetic phase. This result should give insight into the dynamics of the conversion process between charge-order and ferromagnetism with the application and removal of magnetic field.

## IV. Conclusions

In summary, we have examined the relaxation properties of a series of phase separated manganite materials in which the relative phase fraction was varied through sintering of the samples. We find that two different relaxational processes are important in these systems corresponding to either the relaxation of moment orientation or the relaxation of phase fraction. This latter process is unique to the phase-separated nature of the manganites, and may constitute a new sort of magnetic glassiness – based on the macroscopic yet dynamic coexistence of two dissimilar phases. This would add an additional aspect to the many unique properties of the perovskite manganites, but further studies of the detailed dynamics of phase separation, through microscopic probes or theoretical simulations, will be needed to fully characterize the nature of these processes.

**Acknowledgements:** The authors gratefully acknowledge helpful comments from P. Levy and the support of NSF grant DMR-0101318. BGK and SWC were supported by NSF-DMR- 0080008.

**Figure captions**

Figure 1. Average grain size of the polycrystalline samples (from scanning electron microscopy) and the core volume fraction of ferromagnetic phase at 5 K (from $M(H)$ data as described in the text).

Figure 2. Temperature dependence of the resistivity on cooling in zero magnetic field (solid symbols) and in H = 10 T (open symbols) of the entire series of samples.

Figure 3. Magnetization as a function of temperature measured on cooling of the entire series of samples.

Figure 4. Magnetization versus field (after zero field cooling), measured at various temperatures for sample 1100.

Figure 5a. Profiles of the magnetization relaxation for sample 1000 for $H_a$ = 0.05 T and a range of temperatures.

Figure 5b. Profiles of the magnetization relaxation for sample 1000 at 130 K for different applied magnetic fields ($H_a$).

Figure 6. Temperature dependence of the magnetic viscosity for sample 1100 in different applied magnetic fields ($H_a$). Note the two different characteristic behaviors at low and high fields as designated by the open and solid symbols respectively.

Figure 7. The field dependence of the magnetic viscosity at different temperatures for the sample 1100.

Figure 8. The temperature dependence of the magnetic viscosity *S* for all the investigated samples for a 0.05 T applied magnetic field.

Figure 9. The temperature dependence of the magnetic viscosity for all the investigated samples for a 1T applied magnetic field.

Figure 10. The temperature dependence of the magnetic viscosity *S* for the samples 1100 and 1300 for $H_a$ = 2 t and 4 T.

Figure 11. Magnetic viscosity as a function of the sample sintering temperatures, measured at different temperatures and $H_a$ = 1 T.

## References

1. For reviews see: A. P. Ramirez, J. Phys. Condens. Matter **9**, 8171 (1997) and J. M. D. Coey, M. Viret, and S. von Molnar, Adv. Physics **48**,167 (1999).

2. E. Dagotto, T. Hotta, and A. Moreo, Physics Reports **344**, 1 (2001); A. Moreo, S. Yunoki, E. Dagotto, Science **283**, 2034 (1999) and references therein.

3. A. Moreo, M. Mayr, A. Feiguin, S. Yunoki, and E. Dagotto, Physical Review Letters **84**, 5568 (2000).

4. D. Khomskii, Physica B **280**, 325 (2000).

5. P. Schlottmann, Physical Review B **59**, 11484 (1999).

6. S. Okamoto , S. Ishihara, and S. Maekawa, Physical Review B **61**, 451 (2000).

7. L. M. Fisher, A. V. Kalinov, S. E. Savel'ev, I. F. Voloshin, and A. M. Balbashov, J. Phys.: Condens. Matter **10**, 9769 (1998).

8. M. Roy, J. F. Mitchell, and P. Schiffer, Journal of Applied Physics **87** 5831 (2000).

9. S. Balevicius, B. Vengalis, F. Anisimovas, J. Novickij, R. Tolutis, O. Kiprianovic, V. Pyragas, E. E. Tornau, Journal of Magnetism and Magnetic Materials **211**, 243 Sp. Iss. (2000).

10. R. von Helmolt, J. Wecker, T. Lorenz, and K. Samwer, Applied Physics Letters **67**, 2093 (1995).

11. T. Kimura, Y. Tomioka, R. Kumai, Y. Okimoto, and Y. Tokura, Physical Review Letters **83** 3940 (1999).

12. I. F. Voloshin, A. V. Kalinov, S. E. Savel'ev, L. M. Fisher, N. A. Babushkina, L. M. Belova, D. I. Khomskii, K. I. Kugel', JETP Letters **71**, 106 (2000).

13. D. N. H. Nam, R. Mathieu, P. Nordblad, N. V. Khiem, and N. X. Phuc, Physical Review B **62**, 1027 (2000).

14. M. Sirena, L. B. Steren, and J. Guimpel, Physical Review B **64**, 104409 (2001).

15. J. López, P. N. Lisboa-Filho, W. A. C. Passos, W. A. Ortiz, F. M. Aruajo-Moreira, O. F. de Lima, D. Schaniel, and K. Ghosh, Phys. Rev. B **63**, 224422 (2001).

16. V. N. Smolyaninova, C. R. Galley, and R. L. Greene, cond-mat/990708 (1999).

17. P. Levy et al. preprint (cond-mat/0202261).

18. J. M. De Teresa, M. R. Ibarra, J. García, J. Blasco, C. Ritter, P. A. Algarabel, C. Marquina, and A. del Moral, Phys. Rev. Lett. **76**, 3392 (1996).

19. A. Maignan, C. Martin, G. Van Tendeloo, M. Hervieu and B. Raveau, Phys. Rev. B **60**, 15214 (1999).

20. R. Mathieu, P. Nordblad, D. N. H. Nam, N. X. Phuc, N. V. Khiem, Phys. Rev. B **63**, 174405 (2001).

21. A. Maignan, C. Martin, F. Damay, and B. Raveau, J. Hejtmanek, Phys. Rev. B **58**, 2758 (1999).

22. R. S. Freitas, L. Ghivelde, F. Damay, F. Dias, L. F. Cohen, Phys. Rev. B **64**, 144404 (2001).

23. I. G. Deac, J. F. Mitchell and P. Schiffer, Phys. Rev. B **63**, 172408 (2001).

24. See for example M. Roy, J. F. Mitchell, A. P. Ramirez and P. Schiffer, Phys. Rev. B **58**, 5185 (1998) and Y. Tomioka and Y. Tokura in Colossal Magnetoresistance: Charge Ordering and Related Phenomena of Manganese Oxides, C. N. R. Rao and B. Raveau eds., World Scientific, Singapore (1998).

25. F. Parisi, P. Levy, L. Ghivelder, G. Polla, and D. Vega, Phys. Rev. B **63**, 144419 (2001).

26. A. Aharoni, Phys. Rev. B **46**, 5434 (1992).

27. A. Labarta, O. Iglesias, Ll. Balcells, and F. Badia, Phys. Rev. B **48**, 10240 (1993).

28. M. Uehara, S. Mori, C. H. Chen, and S.-W. Cheong, Nature **399**, 560 (1999).

29. K. H. Kim, M. Uehara, C. Hess, P. A. Sharma, and S.-W. Cheong, Phys. Rev. Lett. **84**, 2961 (2000).

30. A. Yakubovskii, K. Kumagai, Y. Furukawa, N. Babushkina, A. Taldenkov, A. Kaul, and O. Gorbenko, Phys. Rev. B **62**, 5337 (200).

31. N. A. Babushkina, L. M. Belova, O. Y.Gorbenko, A. R. Kaul, A. A. Bosak, V. I. Ozhogin, K. I. Kugel, Nature **391**, 159 (1998).

32. V. Kiryukhin, B. G. Kim, V. Podzorov, S.-W. Cheong, T. Y. Koo, J. P. Hill, I. Moon, and Y. H. Jeong, Phys. Rev. B **63**, 024420 (1999).

33. R. Mahesh, R. Mahendiran, A. K. Raychaudhuri, and C. N. R. Rao, Appl. Phys. Lett. **66**, 2291 (1996).

34. V. Podzorov, B. G. Kim, V. Kiryukhin, M. E. Gershenson , and S. -W. Cheong, Phys. Rev. B **64**, 140406 (2001).

35. N. A. Babushkina, L. M. Belova, D. I. Khomskii, K. I. Kugel, O. Yu. Gorbenko, and A. R. Kaul, Phys. Rev. B **59**, 6994 (1999).

36. I. Webman, J. Jortner, and M. H. Cohen, Phys. Rev. B **14**, 4737 (1976).

37. P. Levy, F. Parisi, G. Polla, D. Vega, G. Layva, H, Lanza, R. S. Freitas, and L. Ghivelder, Phys. Rev. B **62**, 6437 (2000).

Fig. 1 Deac et al.

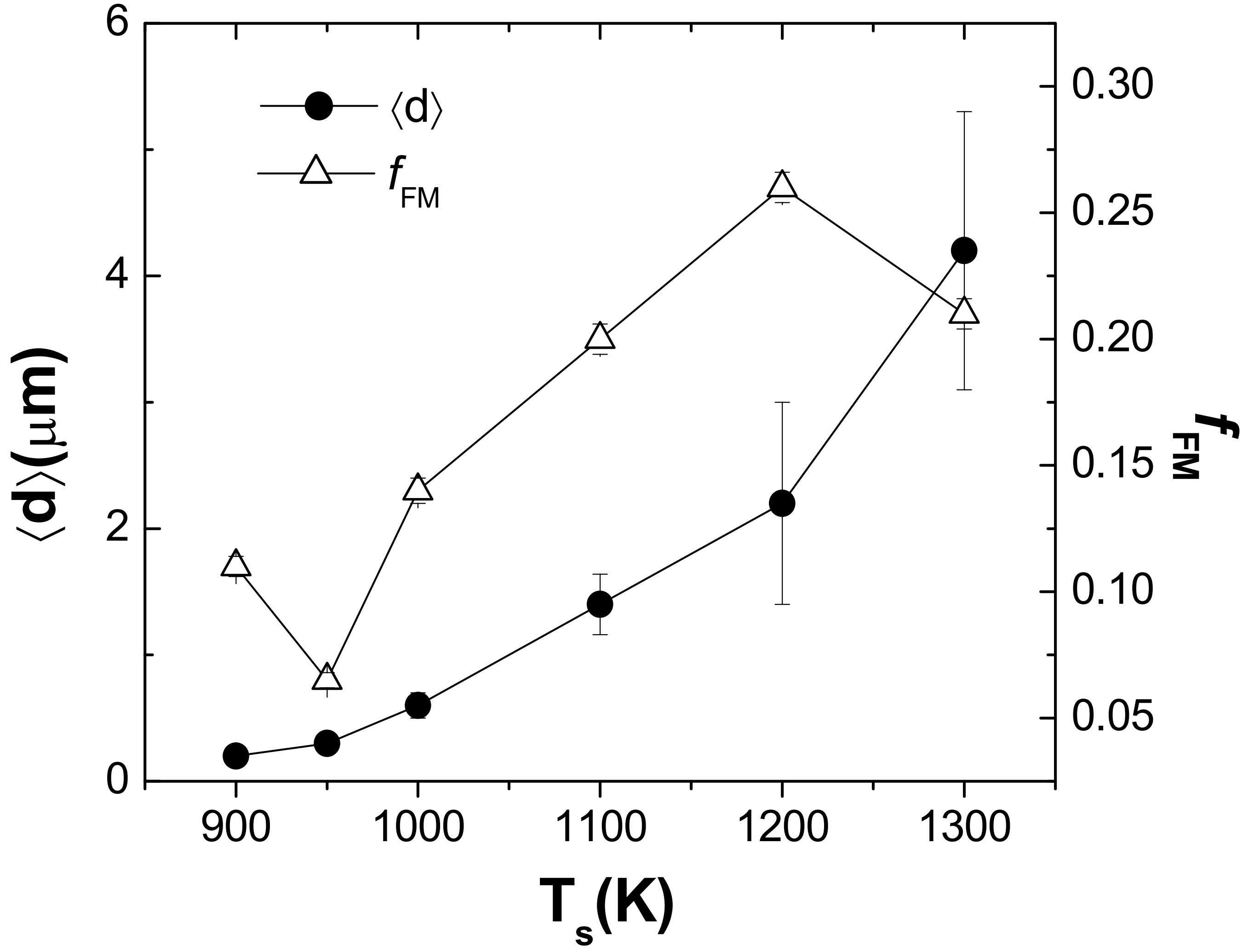

Fig. 2 Deac et al.

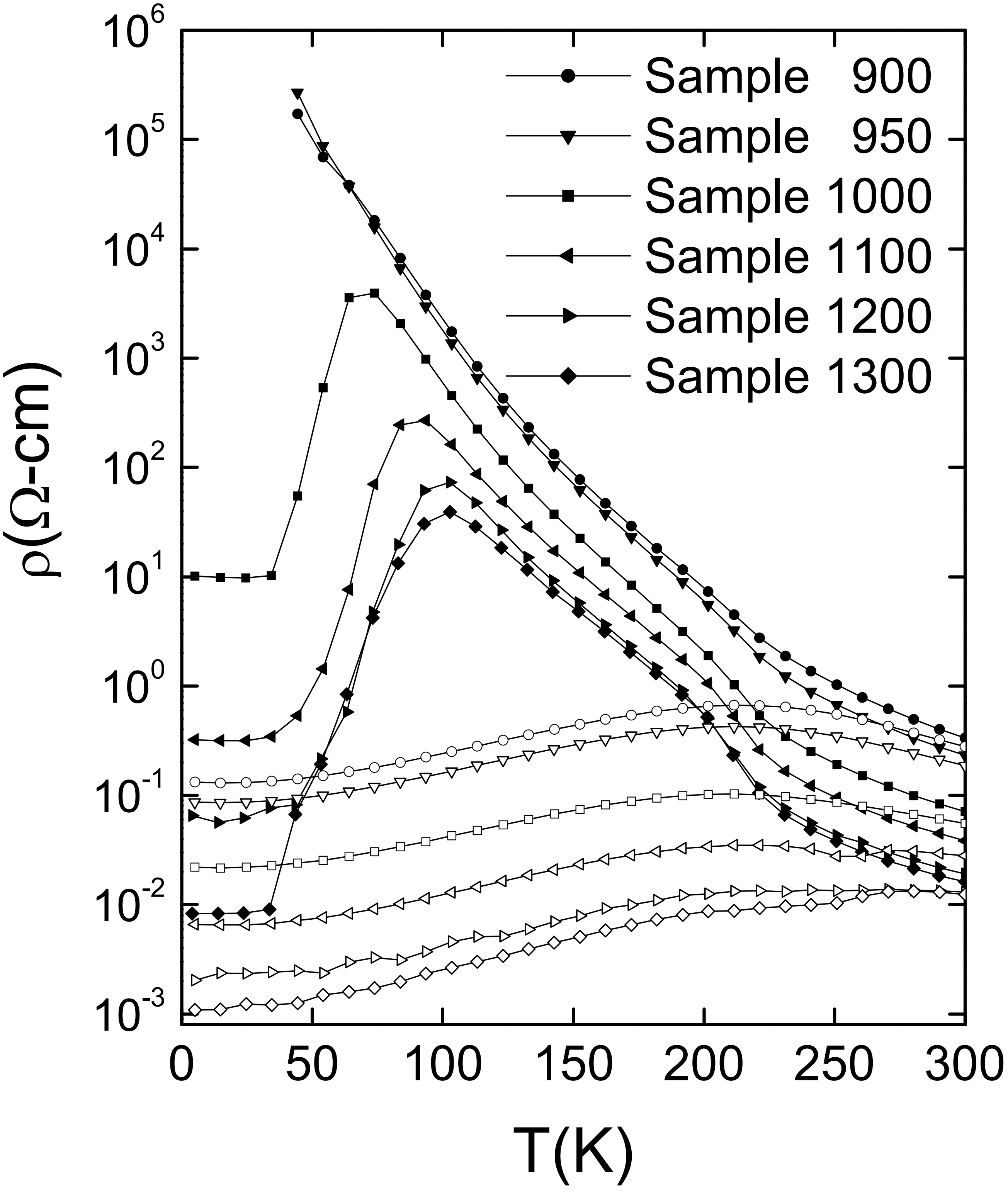

Fig. 3 Deac et al.

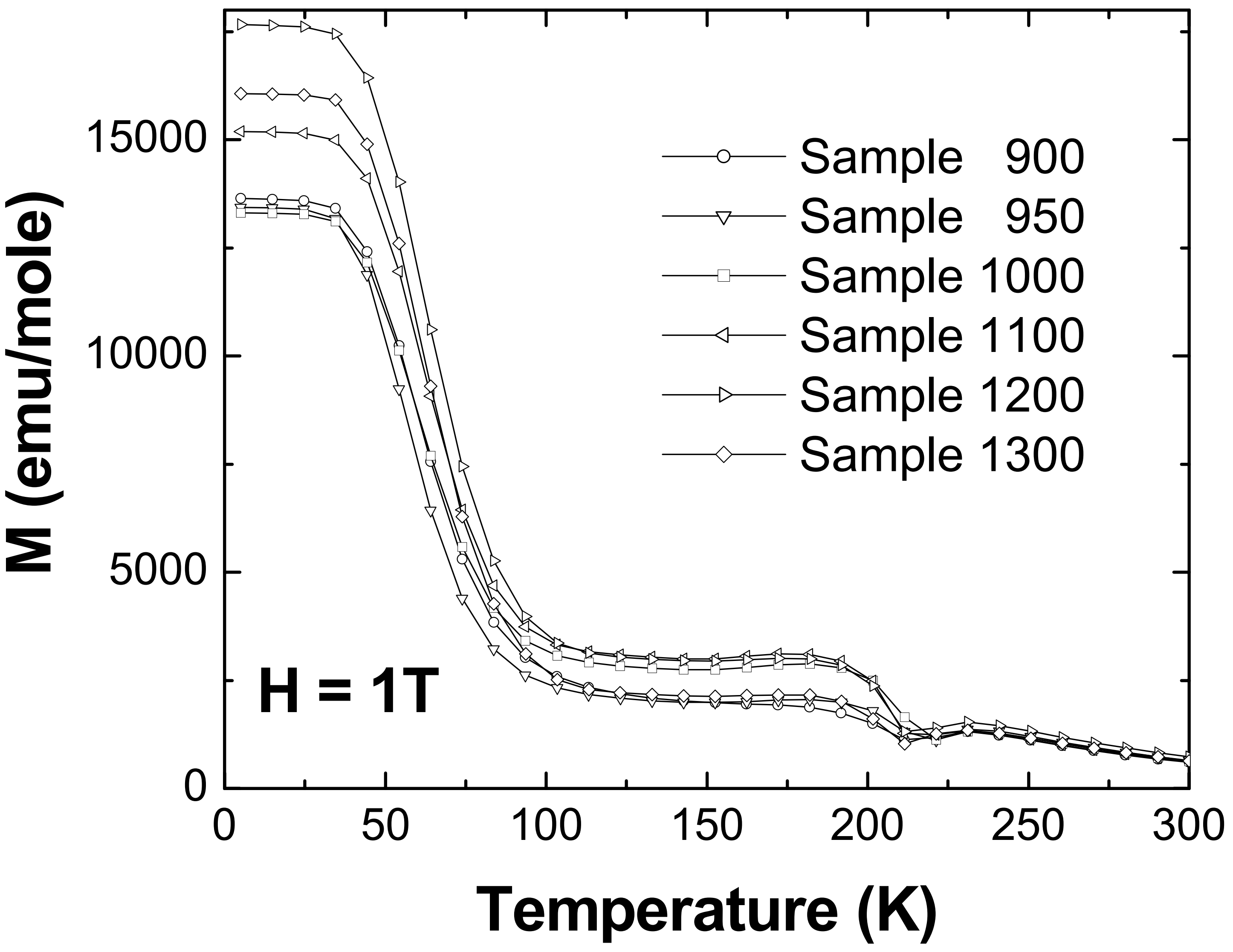

Fig. 4 Deac et al.

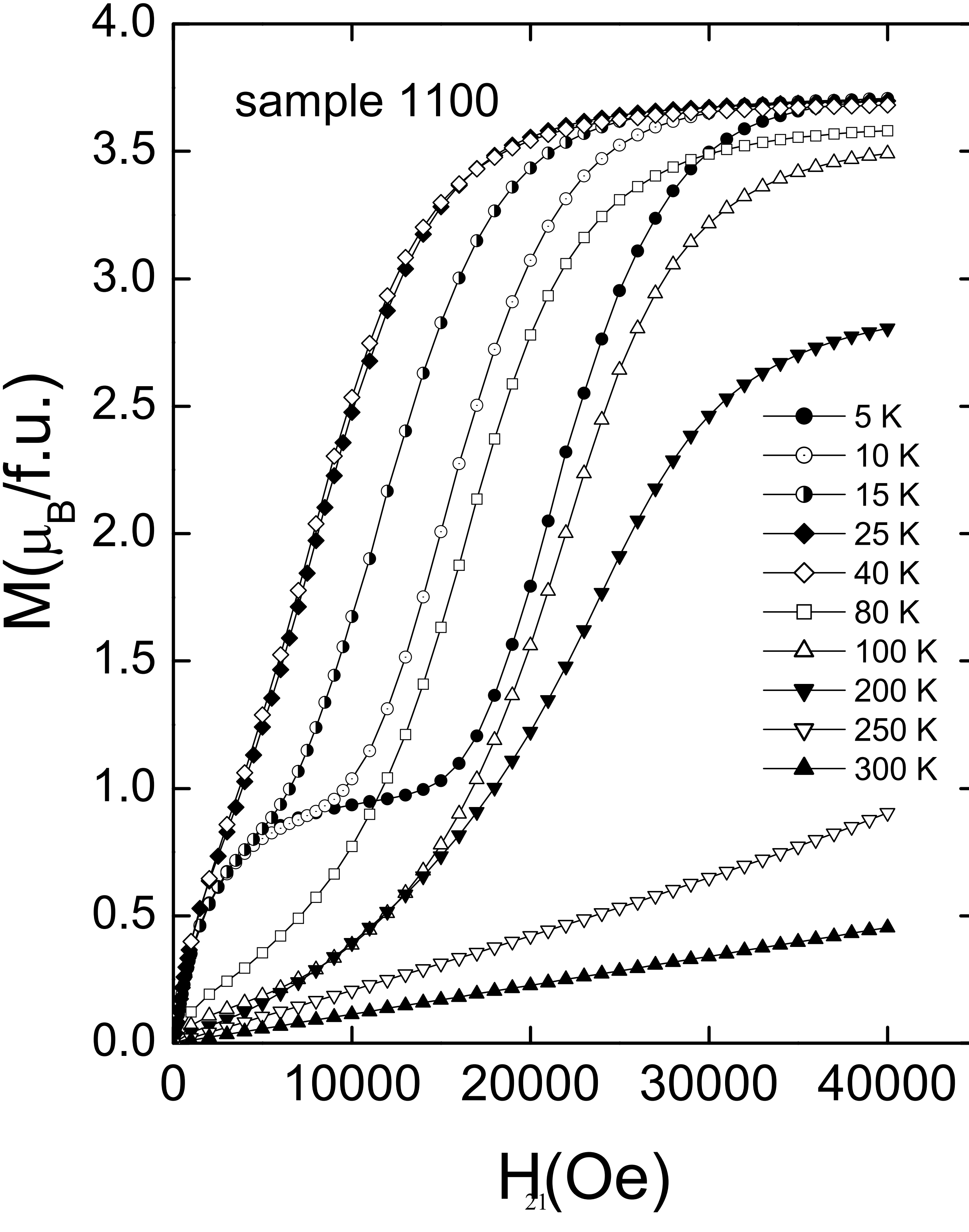

Fig. 5a  Deac et al.

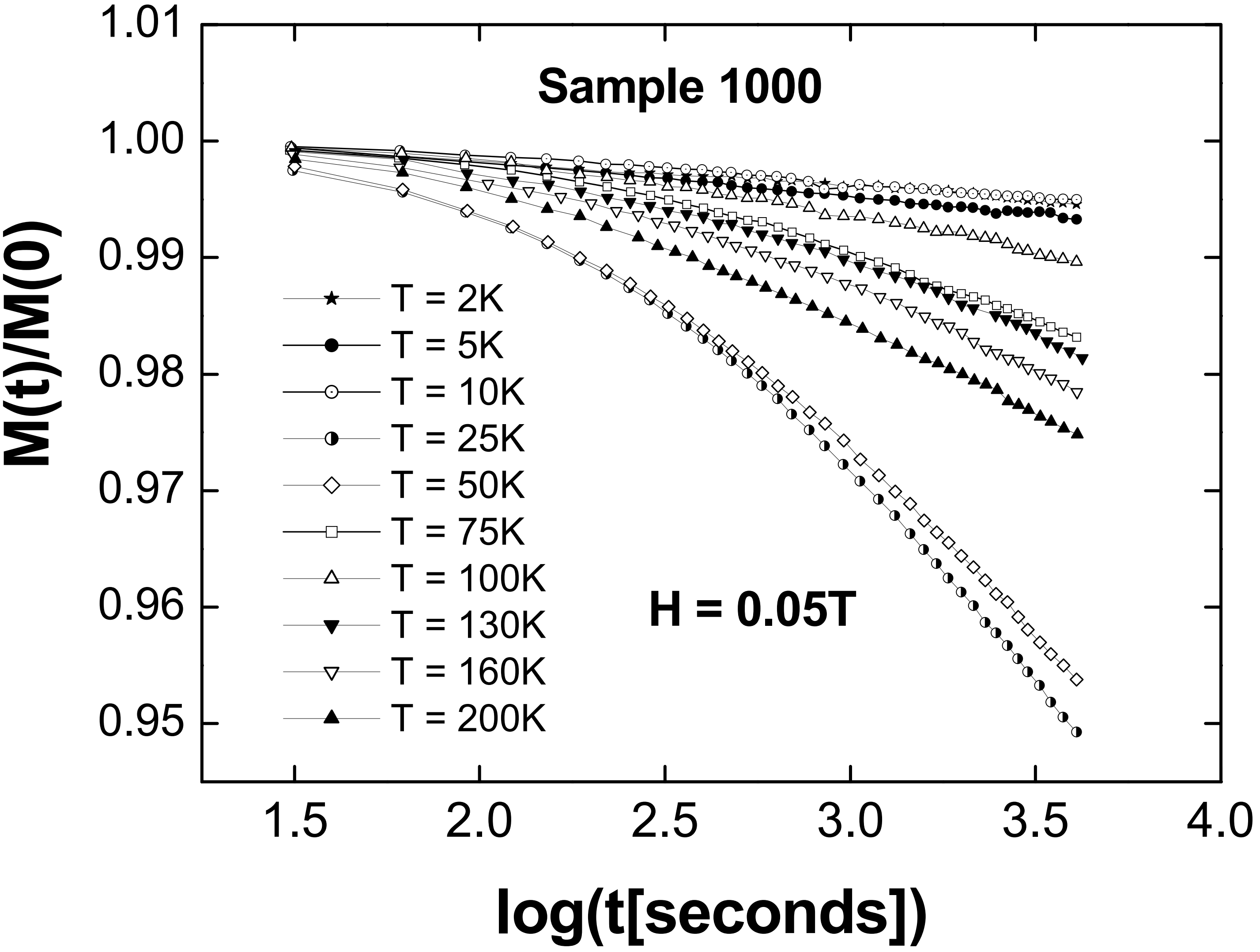

Fig. 5b  Deac et al.

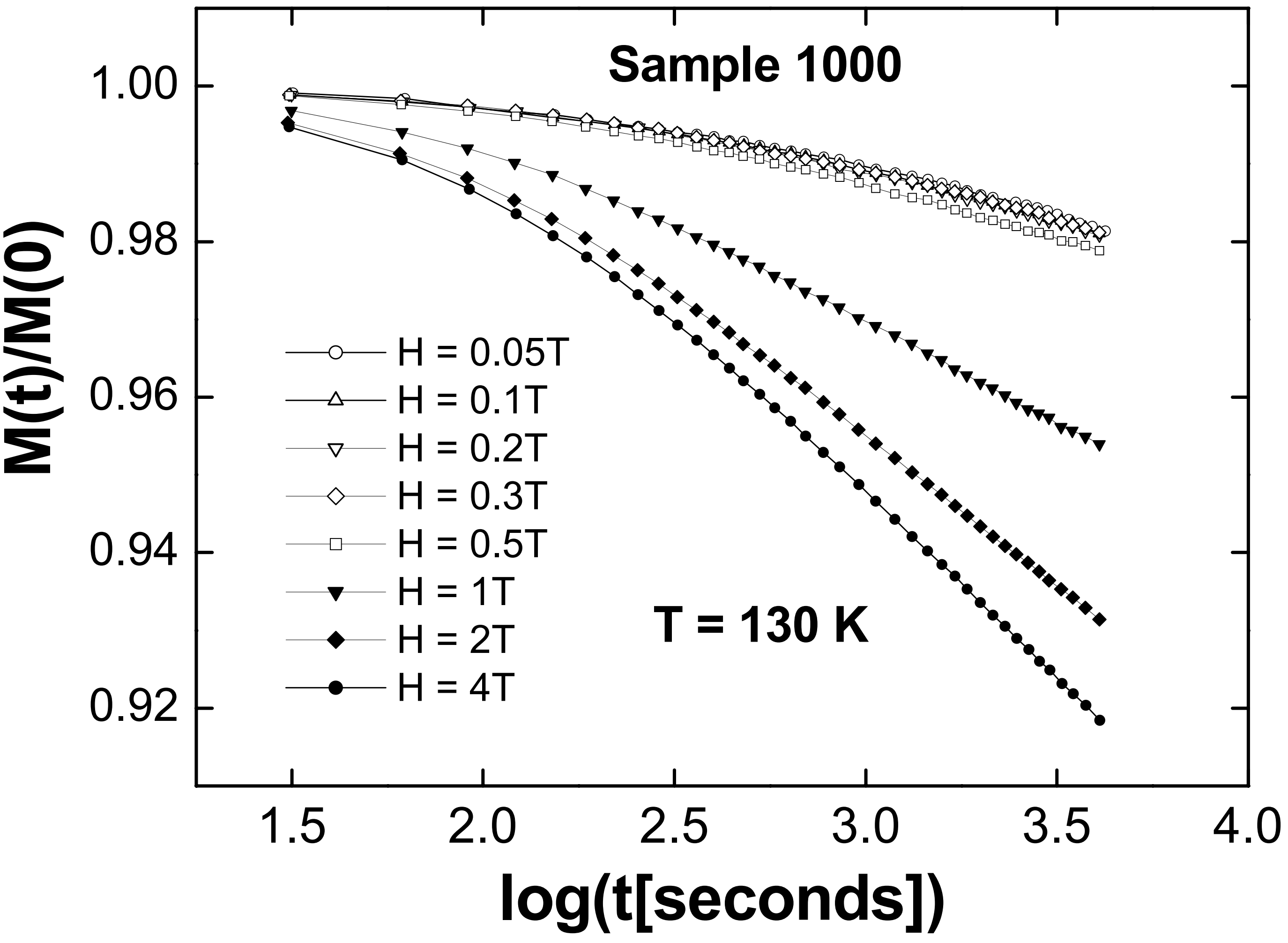

Fig. 6  Deac et al.

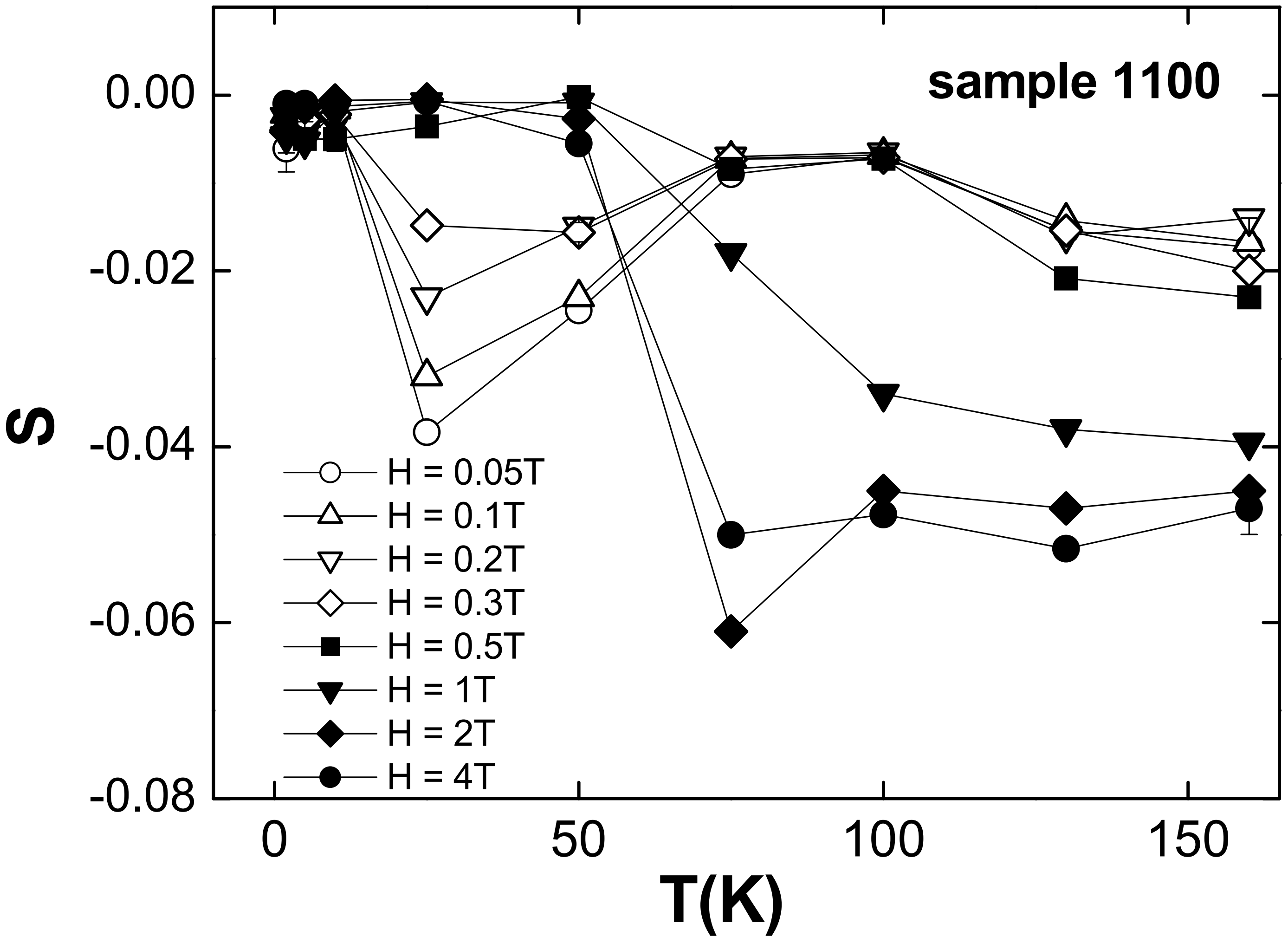

Fig. 7  Deac et al.

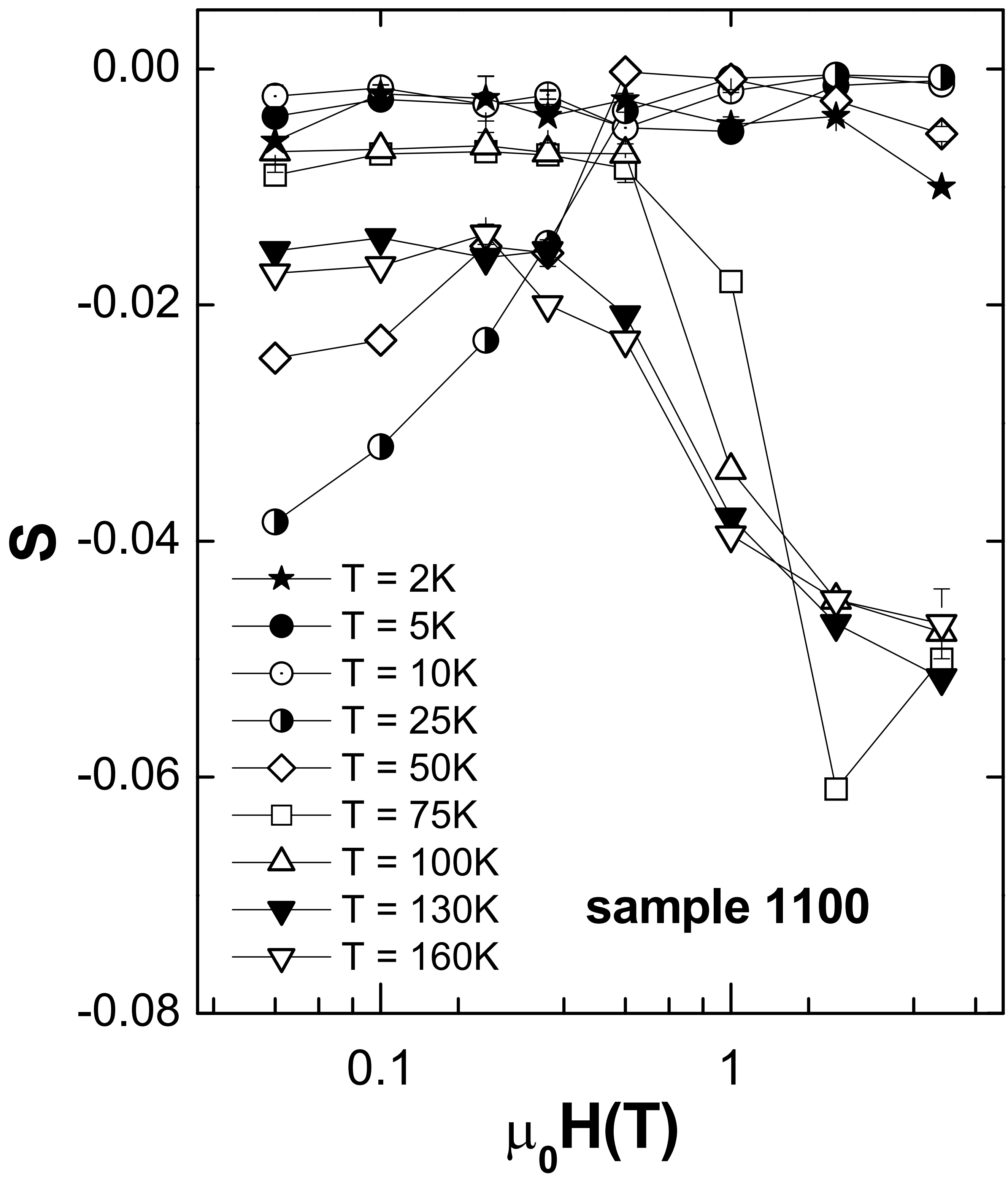

Fig. 8 Deac et al.

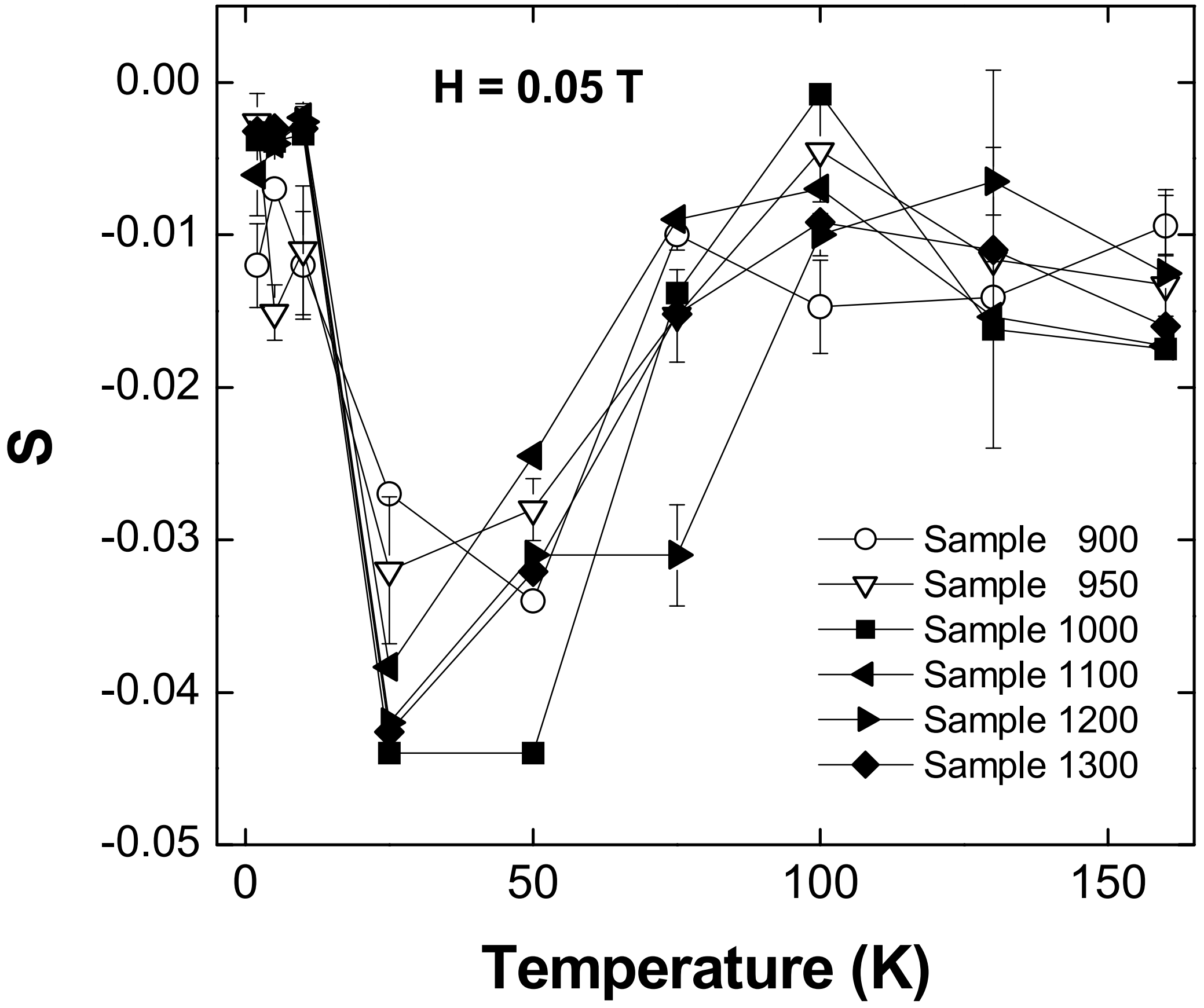

Fig. 9 Deac et al.

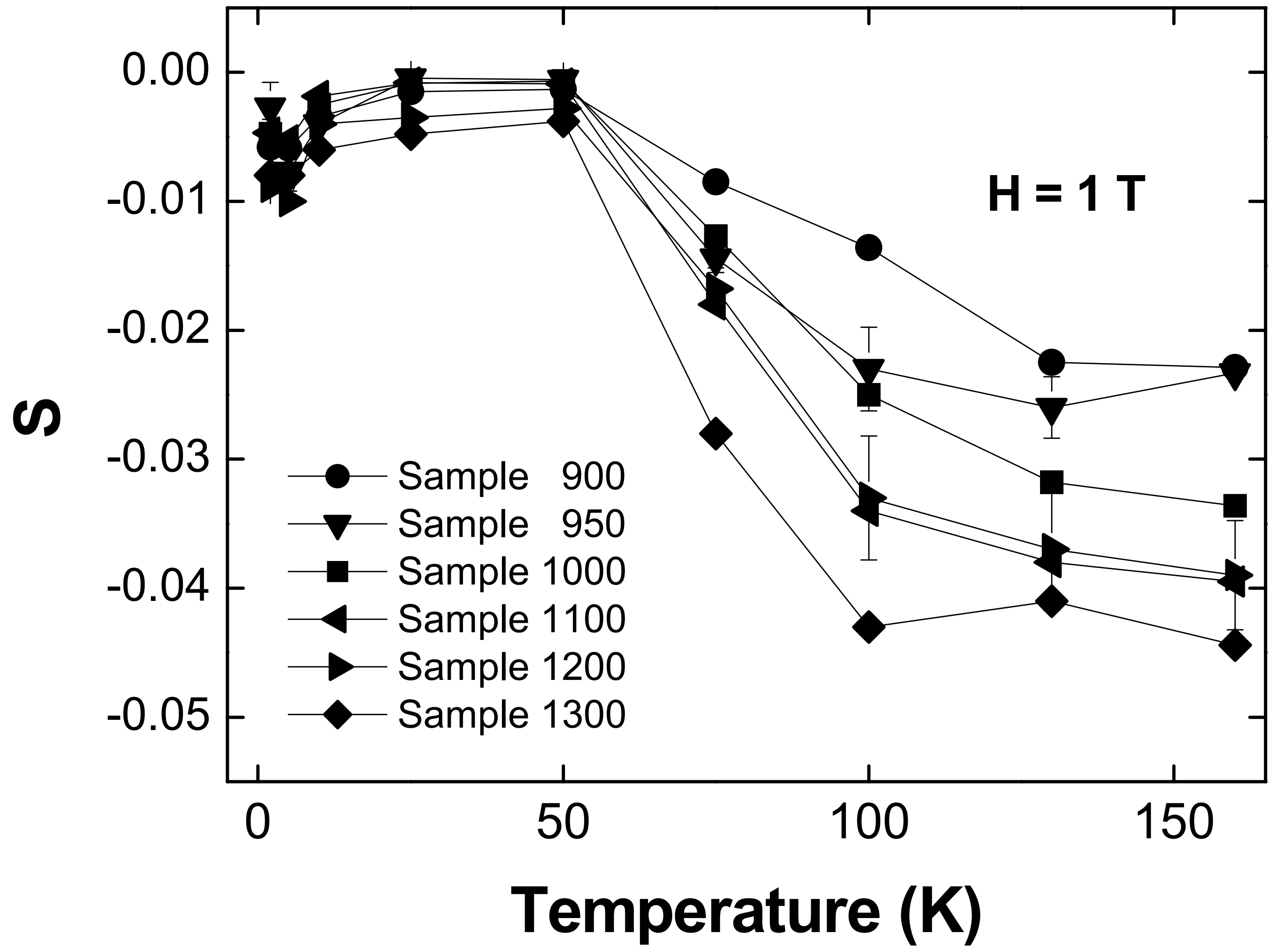

Fig. 10  Deac et al.

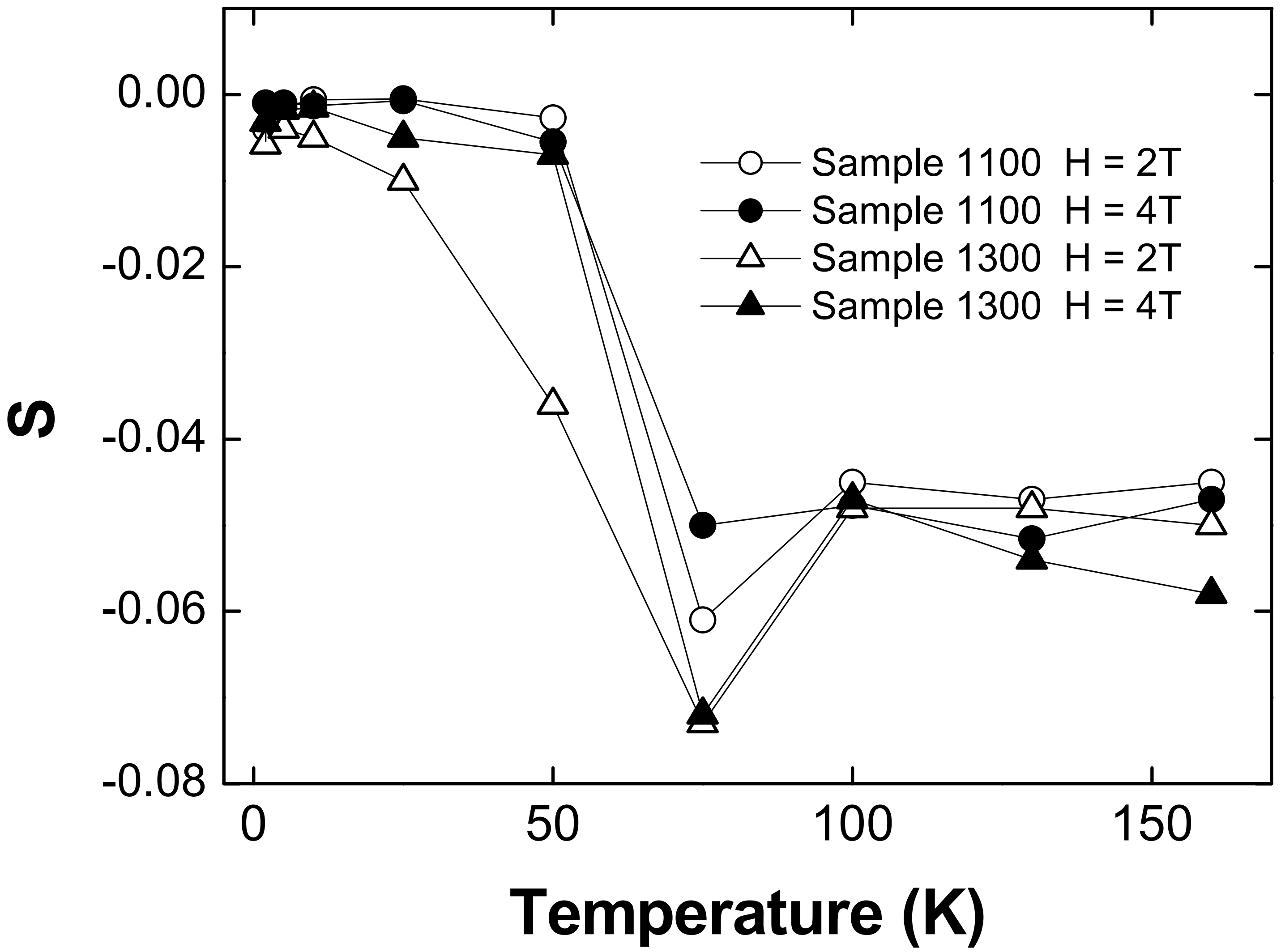

Fig. 11 Deac et al.

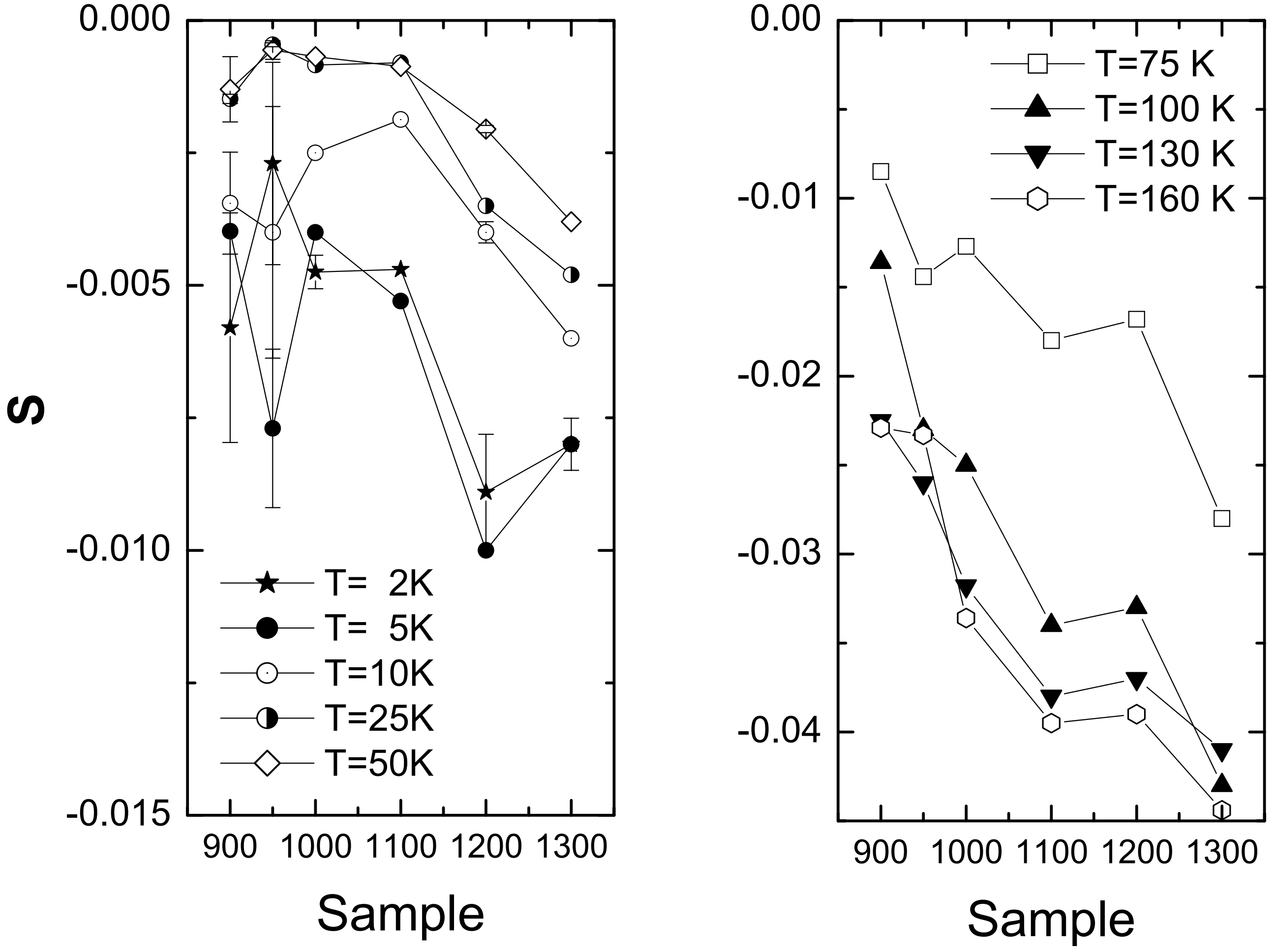